\documentclass{ws-procs9x6}

\begin{document}

\title{Meson elastic and transition form factors}

\author{P.~Maris}

\address{Dept. of Physics, North Carolina State University,\\
Box 8202,  Raleigh,  NC 27695-8202, USA\\
E-mail: pmaris@unity.ncsu.edu}

%%%%%%%%%%%%%%%%%%%%%%%%%%%%%%%%%%%%%%%%%%%%%%%%%%%%%%%%%%%%%%
% You may repeat \author \address as often as necessary      %
%%%%%%%%%%%%%%%%%%%%%%%%%%%%%%%%%%%%%%%%%%%%%%%%%%%%%%%%%%%%%%

\maketitle

\abstracts{The Dyson--Schwinger equations of QCD, truncated to
ladder-rainbow level, are used to calculate meson form factors in
impulse approximation.  The infrared strength of the ladder-rainbow
kernel is described by two parameters fitted to the chiral condensate
and $f_\pi$; the ultraviolet behavior is fixed by the QCD running
coupling.  This obtained elastic form factors $F_\pi(Q^2)$ and
$F_K(Q^2)$ agree well with the available data.  We also calculate the
$\rho \to \pi \gamma$ and $K^\star \to K \gamma$ transition form
factors, which are useful for meson-exchange models.}

%%%%%%%%%%%%%%%%%%%%%%%%%%%%%%%%%%%%%%%%%%%%%%%%%%%%%%%%%%%%%%%%%%%%%%%%%%%%%
\section{Dyson--Schwinger equations}
The set of Dyson--Schwinger equations [DSEs] form a useful tool to
obtain a microscopic description of hadronic properties\cite{review}.
Here we use the DSEs to calculate elastic and transition form factors
of the light mesons.  The dressed quark propagator, as obtained from
its DSE, together with the meson Bethe--Salpeter amplitude [BSA] and
the dressed quark-photon vertex, form the necessary elements for
calculations of form factors in impulse approximation, such as the
pion elastic form factor\cite{Maris:2000bh}.

The DSE for the renormalized quark propagator in
Euclidean space is\cite{review}
\begin{equation}
\label{gendse}
 S(p)^{-1} = i \, Z_2\, /\!\!\!p + Z_4\,m(\mu) + 
        Z_1 \int\!\!\frac{d^4q}{(2\pi)^4} \,
	g^2 D_{\mu\nu}(p-q) \, \textstyle{\frac{\lambda^i}{2}}
	\gamma_\mu \, S(q) \, \Gamma^i_\nu(q,p) \;,
\end{equation}
where $D_{\mu\nu}(k)$ is the dressed-gluon propagator and
$\Gamma^i_\nu(q;p)$ the dressed-quark-gluon vertex.  The most general
solution of Eq.~(\ref{gendse}) has the form 
\mbox{$S(p)^{-1} = i /\!\!\! p A(p^2) + B(p^2)$} and is renormalized 
at spacelike $\mu^2$ according to \mbox{$A(\mu^2)=1$} and
\mbox{$B(\mu^2)=m(\mu)$} with $m(\mu)$ the current quark mass.

Mesons are described by solutions of the Bethe--Salpeter equation
[BSE]
\begin{equation}
 \Gamma_H(p_+,p_-;Q) = \int\!\!\frac{d^4q}{(2\pi)^4} \, 
        K(p,q;Q) \; S(q_+) \, \Gamma_H(q_+,q_-;Q) \, S(q_-)\, ,
\label{homBSE}
\end{equation}
at discrete values of $Q^2 = -m_H^2$, where $m_H$ is the meson mass.
In this equation, $p_+ = p + \eta Q$ and $p_- = p - (1-\eta) Q$ are
the outgoing and incoming quark momenta respectively, and similarly
for $q_\pm$.  The kernel $K$ is the renormalized, amputated $q\bar q$
scattering kernel that is irreducible with respect to a pair of $q\bar
q$ lines.  Together with the canonical normalization condition for
$q\bar q$ bound states, Eq.~(\ref{homBSE}) completely determines the
bound state BSA $\Gamma_H$.  Different types of mesons, such as
pseudoscalar or vector mesons, are characterized by different
Dirac structures.

The quark-photon vertex, \mbox{$\Gamma_\mu(p_+,p_-;Q)$}, with $Q$ the
photon momentum and $p_\pm$ the quark momenta, is the solution of the
inhomogeneous BSE
\begin{equation}
 \Gamma_\mu(p_+,p_-;Q) = Z_2 \, \gamma_\mu + 
        \int\!\!\frac{d^4q}{(2\pi)^4} \, K(p,q;Q) 
        \;S(q_+) \, \Gamma_\mu(q_+,q_-;Q) \, S(q_-)\, .
\label{verBSE}
\end{equation}
Solutions of the homogeneous version of Eq.~(\ref{verBSE}) define
vector meson bound states at timelike photon momenta
\mbox{$Q^2=-m_{\rm V}^2$}.  It follows that $\Gamma_\mu(p_+,p_-;Q)$ has
poles at these locations\cite{Maris:2000bh,Maris:2000sk}.

To solve the BSE, we use a ladder truncation, 
\begin{equation}
        K(p,q;P) \to
        - \alpha^{\rm eff}\big((p-q)^2\big) \, 
	D^0_{\mu\nu}(p-q) 
        \textstyle{\frac{\lambda^i}{2}}\gamma_\mu \otimes
        \textstyle{\frac{\lambda^i}{2}}\gamma_\nu \, ,
\label{eq:ladder}
\end{equation}
in conjunction with the rainbow truncation for the quark DSE,
Eq.~(\ref{gendse}): \mbox{$\Gamma^i_\nu(q,p) \rightarrow
\gamma_\nu\lambda^i/2$} and \mbox{$Z_1 g^2 D_{\mu \nu}(k) \rightarrow
4\pi\alpha^{\rm eff}(k^2) D^0_{\mu\nu}(k)$}.  Here, $D^0_{\mu\nu}(k)$
is the free gluon propagator in Landau gauge, and $\alpha^{\rm
eff}(k^2)$ the effective quark-quark interaction, which reduces to the
one-loop QCD running coupling $\alpha^{\rm 1-loop}(k^2)$ in the
perturbative region.  For the infrared behavior of the interaction, we
employ an Ansatz\cite{Maris:1997tm,Maris:1999nt} that is sufficiently
strong to produce a realistic value for the chiral condensate of about
$(240\,{\rm GeV})^3$.  The model parameters\cite{Maris:1999nt}, along
with the quark masses, are fitted to give a good description of the
chiral condensate, $m_{\pi/K}$ and $f_{\pi}$.  The obtained results
for the light vector meson masses are within 5\% of their experimental
values, and the vector meson electroweak decay constants are within
9\% of the data\cite{Maris:1999nt}.

This truncation preserves both the vector Ward--Takahashi identity
[WTI] for the $q\bar q\gamma$ vertex and the axial-vector WTI,
independent of the details of the effective interaction.  The latter
ensures the existence of massless pseudoscalar mesons associated with
dynamical chiral symmetry breaking\cite{Maris:1997tm,Maris:1998hd}.
In combination with impulse approximation, the former ensures
electromagnetic current conservation\cite{Maris:2000sk}.

%%%%%%%%%%%%%%%%%%%%%%%%%%%%%%%%%%%%%%%%%%%%%%%%%%%%%%%%%%%%%%%%%%%%%%%%%%%%%
\section{Meson Form Factors}
In impulse approximation, meson form factors are generically described
by
\begin{eqnarray}
  I^{abc}(P,Q,K) &=& N_c \int\!\!\frac{d^4q}{(2\pi)^4} \,
	{\rm Tr}\big[ S^a(q) \, \Gamma^{a\bar{b}}(q,q';P) 
        S^b(q')
\nonumber \\ && 
	\Gamma^{b\bar{c}}(q',q'';Q) \, S^c(q'') \,
	\Gamma^{c\bar{a}}(q'',q;K) \big] \,, 
\label{eq:generictri}
\end{eqnarray}
where $q - q' = P$, $q' - q'' = Q$, $q'' - q = K$, and momentum
conservation dictates $P + Q + K = 0$.  In Eq.~(\ref{eq:generictri}),
$S^i$ is the dressed quark propagator with flavor index $i$, and
$\Gamma^{i\bar{j}}(k,k';P)$ stands for a generic vertex function with
incoming quark flavor $j$ and momentum $k'$, and outgoing quark flavor
$i$ and momentum $k$.  Depending on the specific process under
consideration, this vertex function could be a meson BSA or a
quark-photon vertex.  In the calculations discussed below, the
propagators and the vertices are all obtained as solutions of their
respective DSE in rainbow-ladder truncation, without adjusting any of
the model parameters.

%%%%%%%%%%%%%%%%%%%%%%%%%%%%%%%%%%%%%%%%%%%%%%%%%%%%%%%%%%%%%%%%%%
\subsection{Pion and kaon elastic form factors}
There are two diagrams that contribute to meson electromagnetic form
factors: one with the photon coupled to the quark and one with the
photon coupled to the antiquark respectively.  With photon momentum
$Q$, and incoming and outgoing meson momenta $P \mp Q/2$, we can
define a form factor for each of these diagrams\cite{Maris:2000sk}
\begin{eqnarray}
 	2\,P_\nu\,F_{a\bar{b}\bar{b}}(Q^2) &=&
			I_\nu^{abb}(P-Q/2,Q,-(P+Q/2)) \,.
\label{eq:emff}
\end{eqnarray}
We work in the isospin symmetry limit, and thus 
\mbox{$F_{\pi}(Q^2) = F_{u\bar{u}u}(Q^2)$}.  The $K^+$ and $K^0$
form factors are given by \mbox{$F_{K^+}= \frac{2}{3}F_{u\bar{s}u} + 
\frac{1}{3}F_{u\bar s \bar s}$} and \mbox{$F_{K^0} = 
-\frac{1}{3}F_{d\bar{s}d} + \frac{1}{3}F_{d\bar s\bar s}$} respectively.

\begin{figure}[b]
\parbox{2.2in}{\centerline{\epsfxsize=2.1in\epsfbox{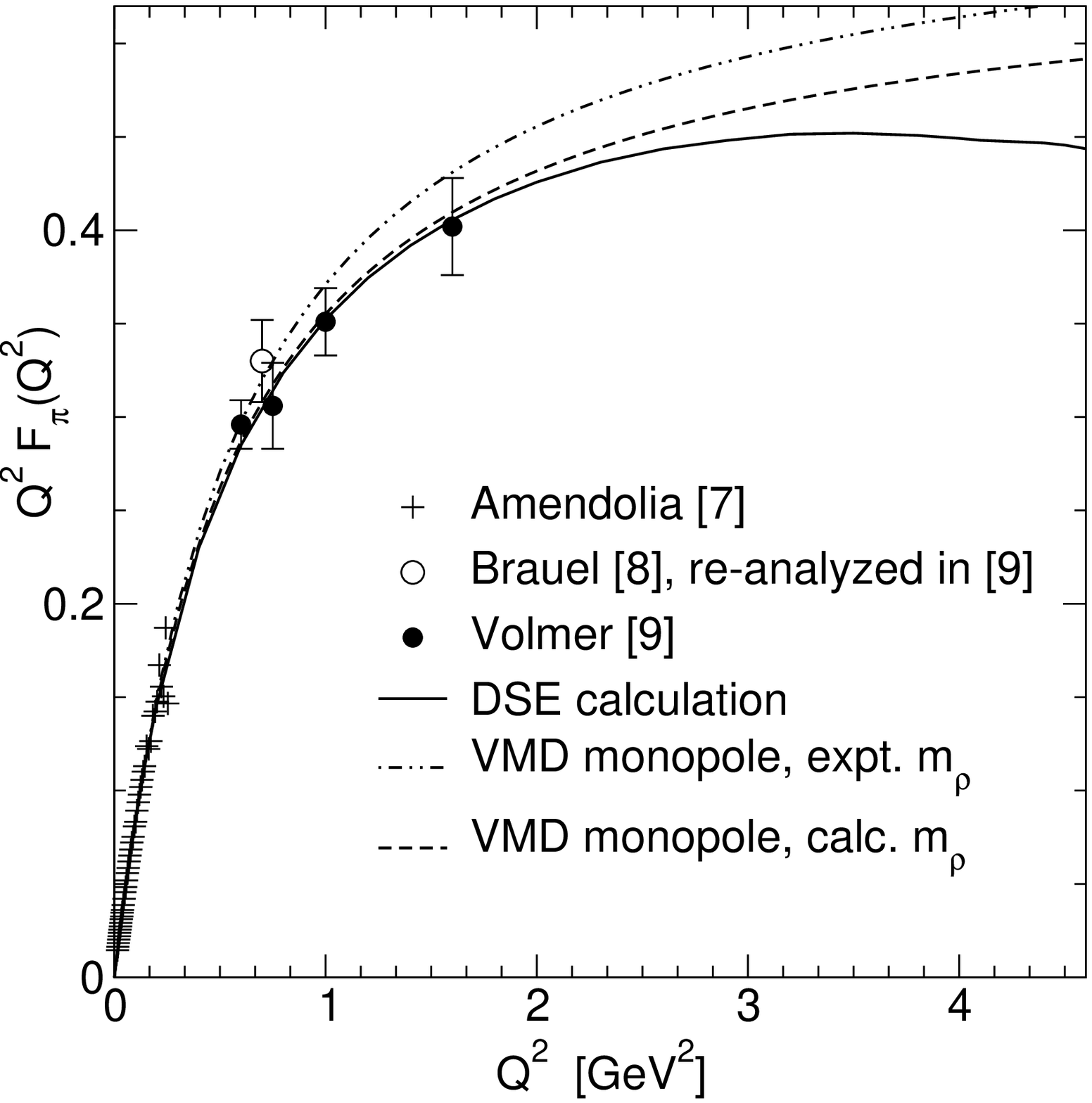}}}
\hfill\parbox{2.2in}{\centerline{\epsfxsize=2.1in\epsfbox{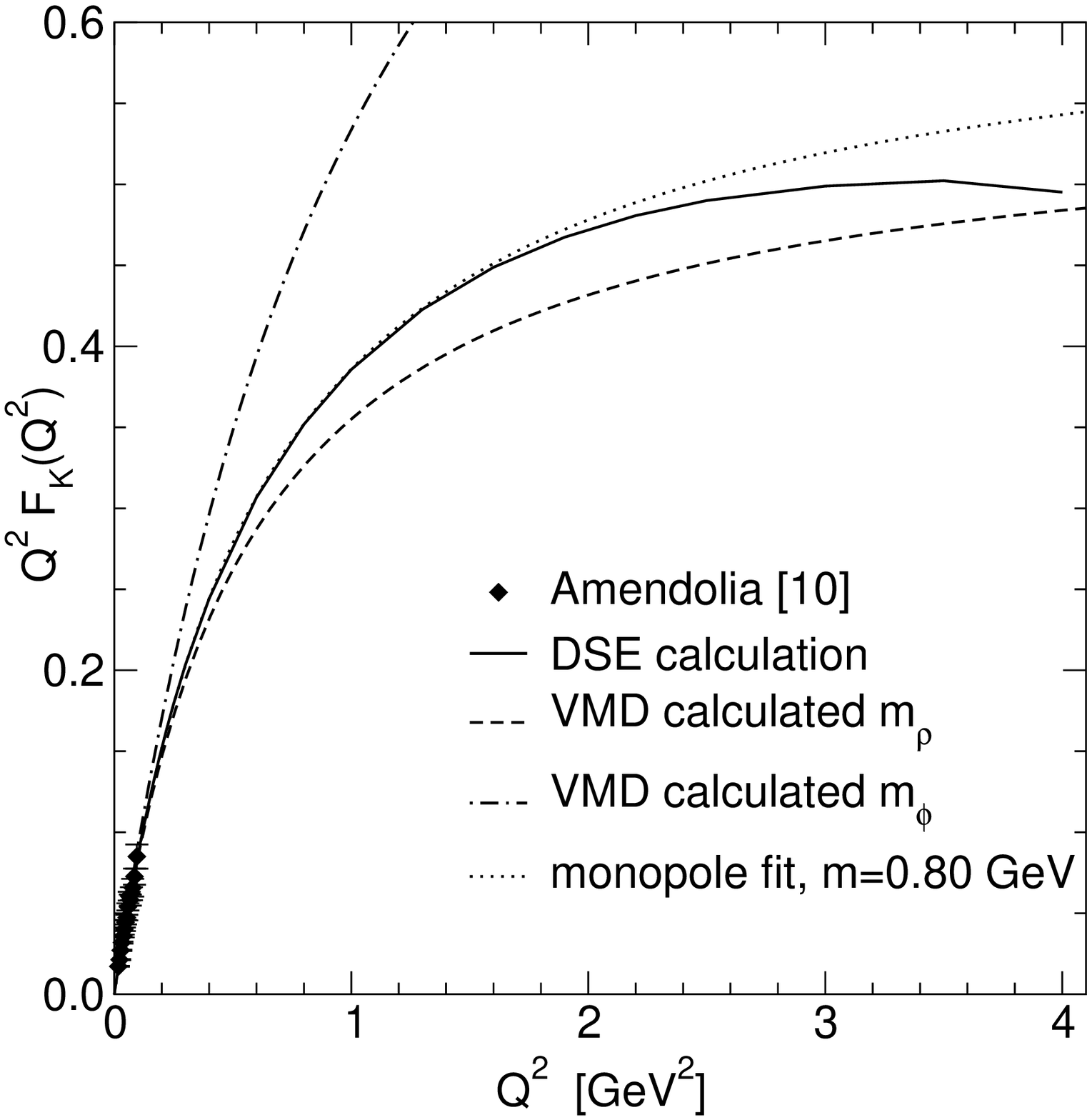}}}
\caption{\label{fig:emff}
On the left, our result for $Q^2 F_\pi(Q^2)$, and right, 
our curve for $Q^2 F_{K^+}(Q^2)$. }
\end{figure}
Our result for $Q^2 F_\pi$ and $F_{K^+}$ are shown in
Fig.~\ref{fig:emff}, together with the experimental
data\cite{A86,desy,Volmer01,A86K}.  Up to about $Q^2 = 2\,{\rm
GeV}^2$, our result for $F_\pi$ can be described very well by a
monopole with our calculated $\rho$-mass, $m_\rho=742\,{\rm MeV}$
(note that our calculated $\rho$-mass is slightly below the
experimental value).  Above this value, our curve starts to deviate
more and more from this naive VMD monopole.  Our result is in
excellent agreement with the most recent JLab data\cite{Volmer01}; it
would be very interesting to compare with future JLab data in the 3 to
5 GeV$^2$ range, where we expect to see a significant deviation from
the naive monopole behavior.

Also our results for $F_K$ agree with the available experimental
data\cite{A86K}, as do both the neutral and the charged kaon charge
radius\cite{Maris:2000sk}.  Again, our curve for $F_K$ can be fitted
quite well by a monopole up to about $Q^2 = 2 \,{\rm GeV}^2$, see
Fig.~\ref{fig:emff}; above 2 GeV$^2$, we see a clear deviation from a
monopole behavior.  Also for this form factor it would be interesting
to compare with future JLab data at larger $Q^2$.

%%%%%%%%%%%%%%%%%%%%%%%%%%%%%%%%%%%%%%%%%%%%%%%%%%%%%%%%%%%%%%%%%%
\subsection{Vector-pseudoscalar-photon transitions}
We can describe the radiative decay of the vector mesons using the
same loop integral, Eq.~(\ref{eq:generictri}), this time with one
vector meson BSA, one pseudoscalar BSA, and one
$q\bar{q}\gamma$-vertex\cite{Maris:2002mz}.  The on-shell
value gives us the coupling constant.  For virtual photons, we can
define a transition form factor $F_{VP\gamma}(Q^2)$, normalized to 1
at $Q^2 = 0$, which can be used in estimating meson-exchange
contributions to hadronic processes\cite{Tandy:1997qf,VODG95}.

In the isospin limit, both the $\rho^0\,\pi^0\,\gamma$ and
$\rho^\pm\,\pi^\pm\,\gamma$ vertices are given by
\begin{eqnarray}
\label{eq:rpgver}
 \frac{1}{3} \, I^{uuu}_{\mu \nu}(P,Q,-(P+Q)) & = &
  \frac{g_{\rho\pi\gamma}}{m_\rho} \; \epsilon_{\mu\nu\alpha\beta} 
	P_\alpha Q_\beta  \, F_{\rho\pi\gamma}(Q^2) \,,
\end{eqnarray}
where $P$ is the $\rho$ momentum.  The $\omega\,\pi\,\gamma$ vertex is
a factor of 3 larger, due to the difference in isospin factors.  For
the $K^\star \to K \gamma$ decay, we have to add two terms: one with
the photon coupled to the $\bar{s}$-quark and one with the photon
coupled to the $u$- or $d$-quark, corresponding to the charged or
neutral $K^\star$ decay respectively\cite{Maris:2002mz}.

As Eq.~(\ref{eq:rpgver}) shows, it is $g_{VP\gamma}/m_V$ that is the
natural outcome of our calculations; therefore, it is this combination
that we give in Table~\ref{tab:vecdec}, together with the
corresponding partial decay widths\cite{Maris:2002mz}.  
\begin{table}[b]
\tbl{Vector meson radiative decays: $g/m$ in GeV$^{-1}$ 
and $\Gamma_{V\to P\gamma}$ in keV.
\label{tab:vecdec}}
{\footnotesize
\begin{tabular}{|l|cc|cc|cc|cc|} \hline
& $g/m$ & $\Gamma_{\rho^\pm\pi^\pm\gamma}$ 	
& $g/m$  & $\Gamma_{\omega\pi\gamma}$	
& $g/m$  & $\Gamma_{K^{\star\pm} K^\pm\gamma}$	
& $g/m$ & $\Gamma_{K^{\star 0} K^0\gamma}$ 	\\\hline
calc.	& 0.69 & 53 & 2.07 & 479 & 0.99	& 90 & 1.19 & 130 \\
expt.\protect\cite{PDG}   
	& 0.74 & 68 & 2.31 & 717 & 0.83	& 50.3 & 1.28 & 116 \\
\hline
\end{tabular}}
\end{table}
The agreement between theory and experiment for $g_{VP\gamma}/m_V$ is
within about 10\%, except for the discrepancy in the charged $K^\star
\to K \gamma$ decay for which we have no explanation.  Likewise the
large difference between the neutral and charged $\rho$ decay width is
beyond the reach of the isospin symmetric impulse approximation.  Note
that part of the difference between the experimental and calculated
decay width comes from the phase space factor because our calculated
vector meson masses deviate up to 5\% from the physical masses.\relax

\begin{figure}[t]
\parbox{2.2in}{\centerline{\epsfxsize=2.1in\epsfbox{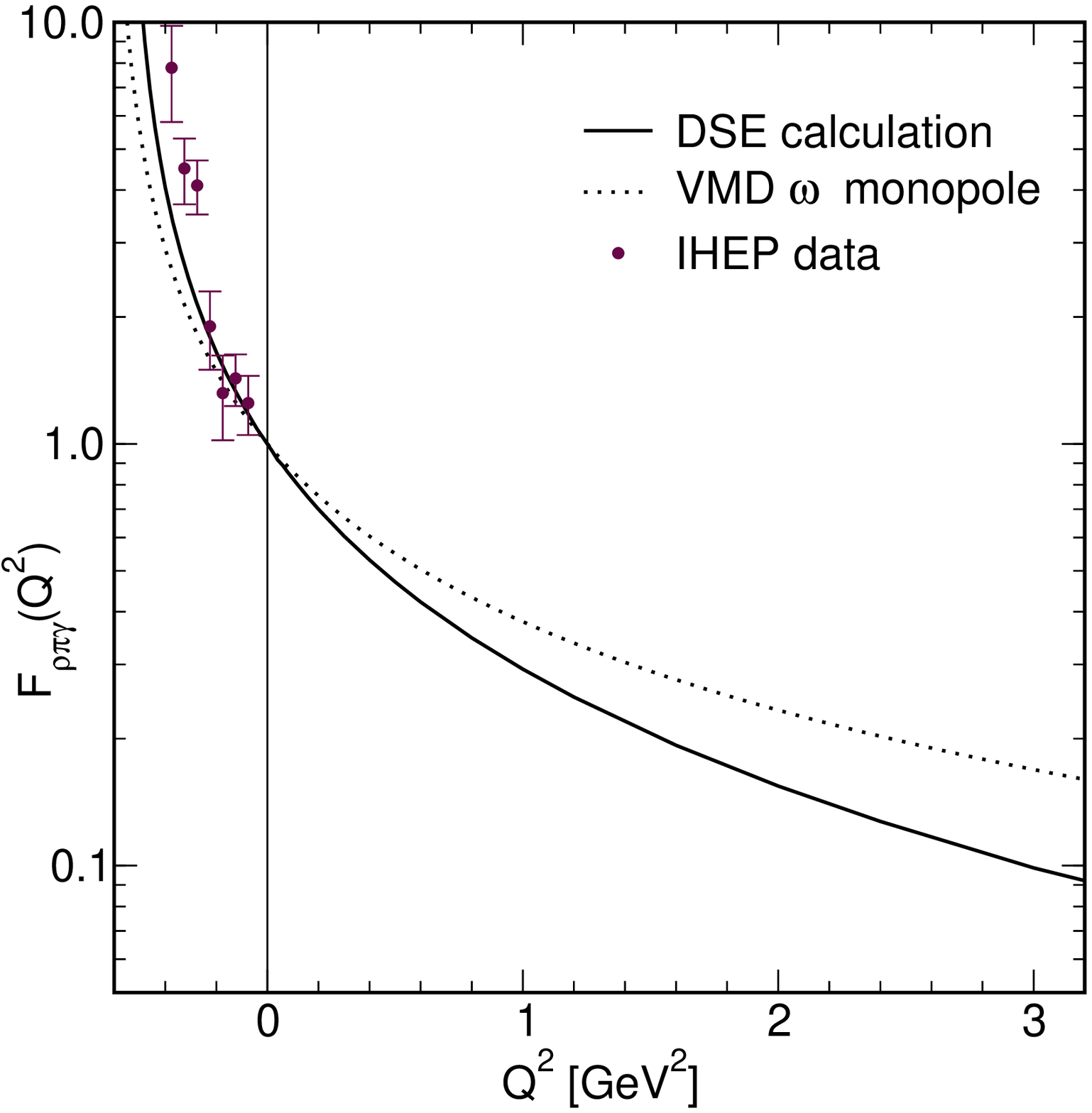}}}
\hfill\parbox{2.2in}{\centerline{\epsfxsize=2.1in\epsfbox{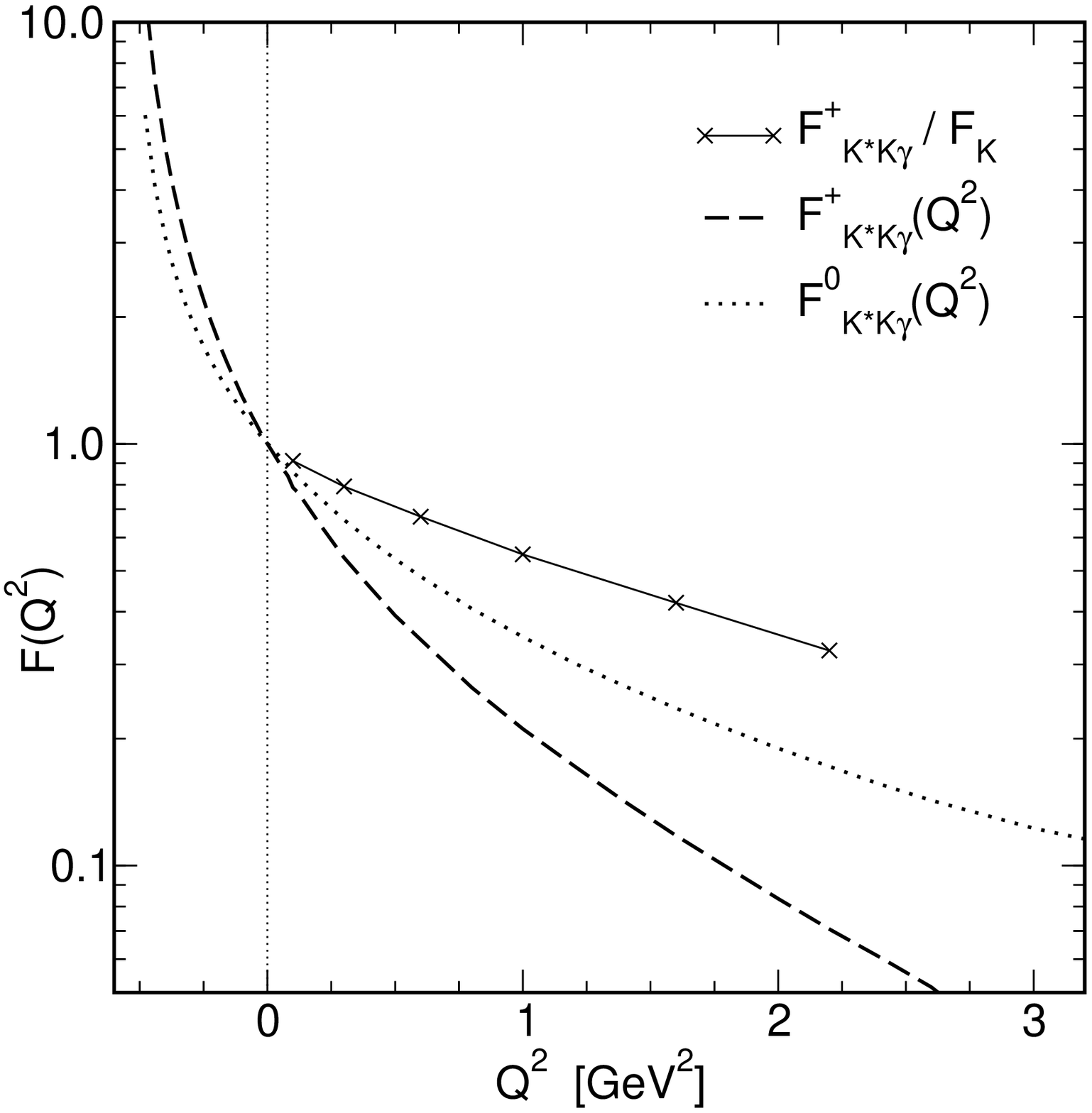}}}
\caption{\label{fig:transff} On the left, our result for
$F_{\omega\pi\gamma}(Q^2)$ with experimental
data\protect\cite{Dzhelyadin:1980tj}, and right, 
our curve for the charged and neutral $F_{K^{\star}K\gamma}(Q^2)$. }
\end{figure}
The corresponding transition form factors are shown in
Fig.~\ref{fig:transff}. In contrast to the elastic form factors, these
transition form factors fall off significantly faster than a VMD-like
monopole; our numerical results\cite{Maris:2002mz} for
$F_{\rho\pi\gamma}(Q^2)$ suggest an asymptotic behavior of $1/Q^4$.
Only in the timelike region, near the vector meson pole,
do we see a clear VMD-like behavior.

For the $K^\star K \gamma$ form factor the situation is more
complicated due to the interference of the diagrams with the photon
coupled to the $s$-quark and to the $u$- or $d$-quark.  This causes
the charged form factor to fall off much more rapidly than both the
neutral $K^\star K \gamma$ form factor and the elastic form factor
$F_K(Q^2)$, as can be seen from Fig.~\ref{fig:transff}.  The latter
implies that the contribution to charged kaon electroproduction from
intermediate $K^\star$ exchange gets suppressed with increasing $Q^2$
compared to the contribution from virtual kaon exchange.  

%%%%%%%%%%%%%%%%%%%%%%%%%%%%%%%%%%%%%%%%%%%%%%%%%%%%%%%%%%%%%%%%%%
\subsection{Remarks on meson electroproduction}
We can use this approach to estimate the range of validity of
meson-dominance models\cite{Maris:2002mz}.  For off-shell pions, the
meson-dominance assumption appears to be quite good for spacelike
momentum transfer of the order of $t \sim 0.1\;{\rm GeV}^2$.  On the
other hand, for heavier mesons such as kaons or $\rho$-mesons, the
naive meson-dominance assumption introduces significant errors, even
at small spacelike values of $t \sim 0.2\;{\rm GeV}^2$.  In addition,
any meson-exchange model to describe meson electroproduction
necessarily introduces off-shell ambiguities.  Clearly, a microscopic
description is needed.

Any microscopic description of meson-electroproduction requires a
quark-gluon description of the nucleon.  It is a difficult task to
combine such a description with the meson form factors considered
here.  Recently, significant progress in describing processes
involving four external particles has been made, using the
rainbow-ladder truncation of the DSEs, in conjunction with an
extention of the impulse approximation\cite{Bicudo:2001jq}.  This
approach incorporates the non-analytic effects of intermediate
meson-exchange contributions while avoiding ambiguous off-shell
definitions.  It has succesfully been applied to $\pi$-$\pi$
scattering, where we can identify the non-analytic contributions from
$\sigma$- and $\rho$-exchange to the S-wave and P-wave scattering
amplitudes respectively\cite{Cotanch:prep}.  Furthermore, it was shown
to reproduce the correct chiral limit\cite{Bicudo:2001jq}.  We plan to
extend this approach to describe processes such as $\gamma \pi\pi\pi$,
pion compton scattering, and meson electroproduction.

%%%%%%%%%%%%%%%%%%%%%%%%%%%%%%%%%%%%%%%%%%%%%%%%%%%%%%%%%%%%%%%%%%%%%%%%%%%%%
\section*{Acknowledgments}
%
%Most of this work was done in collaboration with Peter Tandy; I would
%also like to thank Steve Cotanch, Cheung-Ryong Ji and Craig Roberts
%for useful discussions.  This work was funded by the US Department of
%Energy under grants No. DE-FG02-96ER40947 and DE-FG02-97ER41048, and
%benefitted from the resources of the National Energy Research
%Scientific Computing Center.

Most of this work was done in collaboration with Peter Tandy.  
This work was funded by the US Department of Energy under grants
No. DE-FG02-96ER40947 and DE-FG02-97ER41048, and benefitted from the
resources of the National Energy Research Scientific Computing Center.

%%%%%%%%%%%%%%%%%%%%%%%%%%%%%%%%%%%%%%%%%%%%%%%%%%%%%%%%%%%%%%%%%%%%%%%%%%%%%
%  Start references here:


\begin{thebibliography}{99}
%
\bibitem{review}
C.~D.~Roberts and S.~M.~Schmidt,
%``Dyson-Schwinger equations: Density, temperature and continuum strong  QCD,''
Prog.\ Part.\ Nucl.\ Phys.\  {\bf 45S1}, 1 (2000);
%%CITATION = NUCL-TH 0005064;%%
R.~Alkofer and L.~von Smekal,
%``The infrared behavior of QCD Green's functions: Confinement, dynamical  symmetry breaking, and hadrons as relativistic bound states,''
Phys.\ Rept.\  {\bf 353}, 281 (2001);
%%CITATION = HEP-PH 0007355;%%
%
\bibitem{Maris:2000bh}
P.~Maris and P.~C.~Tandy,
%``The quark photon vertex and the pion charge radius,''
Phys.\ Rev.\ C {\bf 61}, 045202 (2000).
%%CITATION = NUCL-TH 9910033;%%
%
\bibitem{Maris:2000sk}
P.~Maris and P.~C.~Tandy,
%``The pi, K+, and K0 electromagnetic form factors,''
Phys.\ Rev.\ C {\bf 62}, 055204 (2000).
%%CITATION = NUCL-TH 0005015;%%
%
\bibitem{Maris:1997tm}
P.~Maris and C.~D.~Roberts,
%``pi and K meson Bethe-Salpeter amplitudes,''
Phys.\ Rev.\ C {\bf 56}, 3369 (1997).
%%CITATION = NUCL-TH 9708029;%%
%
\bibitem{Maris:1999nt}
P.~Maris and P.~C.~Tandy,
%``Bethe-Salpeter study of vector meson masses and decay constants,''
Phys.\ Rev.\ C {\bf 60}, 055214 (1999).
%%CITATION = NUCL-TH 9905056;%%
%
\bibitem{Maris:1998hd}
P.~Maris, C.~D.~Roberts and P.~C.~Tandy,
%``Pion mass and decay constant,''
Phys.\ Lett.\ B {\bf 420}, 267 (1998).
%%CITATION = NUCL-TH 9707003;%%
%
\bibitem{A86} 
S.~R.~Amendolia {\it et al.}, 
%[NA7 Collaboration],
%``A Measurement Of The Space - Like Pion Electromagnetic Form-Factor,''
Nucl.\ Phys.\ B {\bf 277}, 168 (1986).
%%CITATION = NUPHA,B277,168;%%
%
\bibitem{desy} P. Brauel {\it et al.}, 
Z.\ Phys.\ {\bf C3}, 101 (1979).
%
\bibitem{Volmer01} 
J. Volmer {\it et al.}, 
%[The JLab F(pi) Collaboration],
Phys.\ Rev.\ Lett.\  {\bf 86}, 1713 (2001).
%
\bibitem{A86K} 
S.~R.~Amendolia {\it et al.},
%``A Measurement Of The Kaon Charge Radius,''
Phys.\ Lett.\ B {\bf 178}, 435 (1986).
%%CITATION = PHLTA,B178,435;%%
%
\bibitem{Maris:2002mz}
P.~Maris and P.~C.~Tandy,
%``Electromagnetic transition form factors of light mesons,''
Phys.\ Rev.\ C {\bf 65}, 045211 (2002).
%%CITATION = NUCL-TH 0201017;%%
%
\bibitem{Tandy:1997qf}
P.~C.~Tandy,
%``Hadron physics from the global color model of QCD,''
Prog.\ Part.\ Nucl.\ Phys.\  {\bf 39}, 117 (1997).
%[nucl-th/9705018].
%%CITATION = NUCL-TH 9705018;%%
%
\bibitem{VODG95}
J.~W.~Van Orden, N.~Devine and F.~Gross,
%``Elastic electron scattering from the deuteron using the gross equation,''
Phys.\ Rev.\ Lett.\  {\bf 75}, 4369 (1995).
%%CITATION = PRLTA,75,4369;%%
%
\bibitem{PDG}
Particle Data Group, C.~Caso {\it et al.},  
Eur.\ Phys.\ J.\  {\bf C3}, 1 (1998).
%%CITATION = EPHJA,C3,1;%%
%
\bibitem{Dzhelyadin:1980tj}
R.~I.~Dzhelyadin {\it et al.},
%``Study Of The Electromagnetic Transition Form-Factor In Omega $\to$ Pi0 Mu+ Mu- Decay,''
Phys.\ Lett.\ B {\bf 102}, 296 (1981)
[JETP Lett.\  {\bf 33}, 228 (1981)].
%%CITATION = PHLTA,B102,296;%%
%
\bibitem{Bicudo:2001jq}
P.~Bicudo {\it et al.},
%S.~Cotanch, F.~Llanes-Estrada, P.~Maris, E.~Ribeiro and A.~Szczepaniak,
%``Chirally symmetric quark description of low energy pi pi scattering,''
Phys.\ Rev.\ D {\bf 65}, 076008 (2002).
%%CITATION = HEP-PH 0112015;%%
%
\bibitem{Cotanch:prep}
S.~Cotanch and P.~Maris, in preparation.

\end{thebibliography}
\end{document}